\documentclass[aps,pra,floatfix,reprint]{revtex4-1}
\usepackage{amsmath}
\usepackage{amsfonts}
\usepackage{amssymb}
\usepackage{graphicx}
\usepackage{color}
\usepackage{hyperref}
\usepackage[]{cleveref}

\usepackage{yypreamble}

\begin{document}

\title{Quantum backaction and noise interference in asymmetric two-cavity optomechanical systems}

\author{Yariv Yanay}
\author{Jack C. Sankey}
\author{Aashish A. Clerk}
\affiliation{Department of Physics, McGill University, Montreal, Canada H3A 2T8}

\date{\today}

\begin{abstract}
We study the effect of cavity damping asymmetries on backaction in a ``membrane-in-the-middle'' optomechanical system, where a mechanical mode modulates the coupling between two photonic modes.  
We show that in the adiabatic limit, this system generically realizes a dissipative optomechanical coupling, with an effective position-dependent
photonic damping rate.  The resulting quantum noise interference can be used to ground-state cool a mechanical resonator in the unresolved sideband regime.  We explicitly demonstrate how quantum noise interference controls linear backaction effects, and show that this interference persists even outside the adiabatic limit.  For a one-port cavity in the extreme bad-cavity limit, the interference allows one to cancel all linear backaction effects.  This allows continuous measurements of position-squared, with no stringent constraints on the single-photon optomechanical coupling strength.  In contrast, such a complete cancellation is not possible in the good cavity limit. This places strict bounds on the optomechanical coupling required for quantum non-demolition measurements of mechanical energy, even in a one-port device.   

\end{abstract}


\maketitle

\section{Introduction}

The field of quantum optomechanics has largely focused on a relatively simple system, where photons in a single mode of a resonant cavity interact with the motion of a mechanical resonator.  This kind of system has been at the heart of a number of recent experimental breakthroughs, ranging from the near ground-state cooling of a mechanical mode \cite{Teufel2011,Chan2011,Peterson2016} to the generation of squeezed light \cite{Brooks2012,Safavi-Naeini2013}.  Potentially richer behavior can be realized in a system where a mechanical resonator couples to two optical modes, which are in turn tunnel-coupled to one another \cite{Thompson2008}.  By using the avoided crossing of the optical normal modes, such systems can realize an effective quadratic optomechanical coupling, where the adiabatic cavity normal mode frequencies have no linear dependence on the mechanical position $x$, but instead depend on $x^2$.  Such devices could allow quantum non-demolition (QND) measurement of phonon and photon number \cite{Thompson2008,Jayich2008,Ludwig2012}, and can realize novel kinds of optomechanical cooling and squeezing \cite{Bhattacharya2008,Nunnenkamp2010a,Asjad2014}.  Quadratic couplings were first realized in ``membrane-in-the-middle" systems, where a moveable membrane is placed between two fixed mirrors of a Fabry-Perot cavity \cite{Thompson2008,Sankey2010}, and more recently in several other experimental setups \cite{Paraiso2015,Kaviani2015,Doolin2014,Brawley2016}.

A key issue in such systems is the presence of residual linear backaction, that is, the linear coupling of external noise sources to the mechanics through the cavity. Such backaction hinders QND phonon-number detection. As discussed extensively by Miao et al.~\cite{Miao2009}, linear backaction persists, even though the adiabatic normal-mode frequencies depend on $x^2$, because the adiabatic wavefunctions depend linearly on $x$. When each cavity normal mode is coupled to an independent dissipative reservoir, the linear noise is proportional to the product of both dissipation rates, suggesting that it could be eliminated if one had a truly single-port cavity.

\begin{figure}[t] 
   \centering
   \includegraphics[width=0.9\columnwidth]{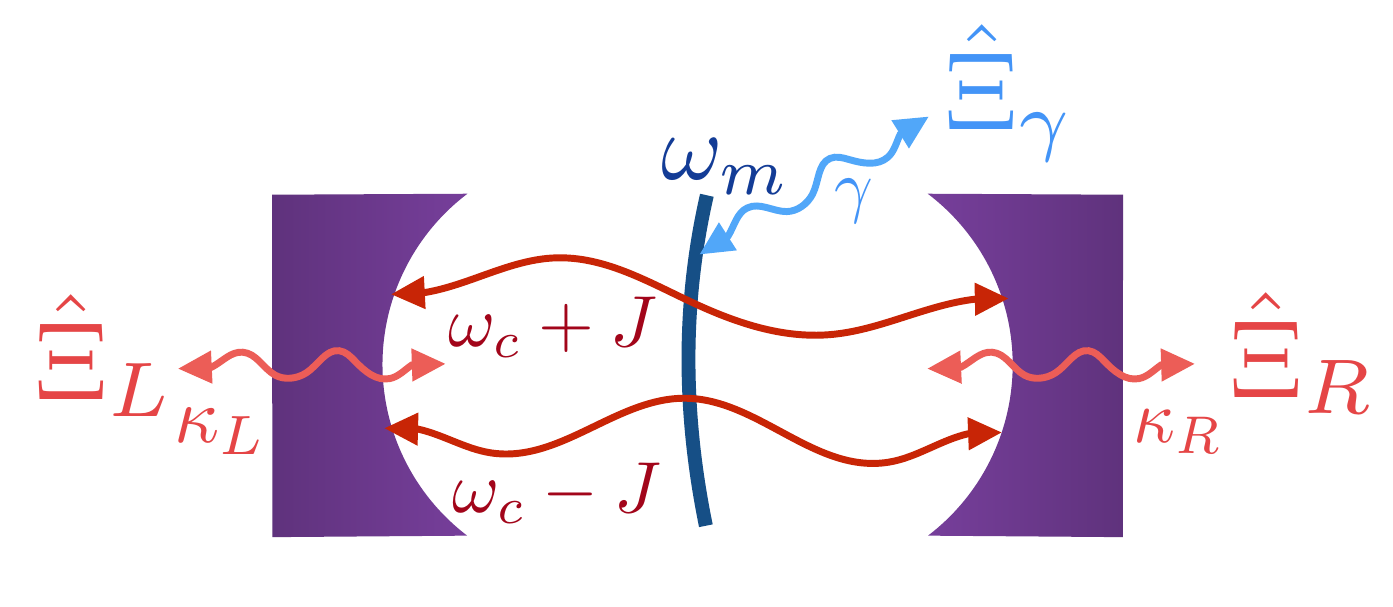} 
   \caption{Schematic of the two-cavity optomechanical system described by \cref{eq:Htot}. Two cavity normal modes, with energies $\omega_{c} \pm J$,    couple to a moving membrane (frequency $\omega_{m}$). The optical normal modes are {\it both} coupled to two independent reservoirs, $L, R$, with different damping rates, while the membrane is coupled to a third reservoir at damping rate $\gamma$. }
   \label{fig:diagram}
\end{figure}

In this work, we revisit linear backaction effects in quadratic-coupling optomechanics, focusing on 
the case of a realistic, asymmetric two-port cavity.  In general, the two dissipative ports each couple to both photonic normal-modes as illustrated in \cref{fig:diagram}, implying that the two modes see correlated dissipation and noise.  In the adiabatic limit, where the the splitting of the photonic modes is much greater than all other energy scales, we show that one obtains an effective dissipative optomechanical coupling \cite{Elste2009}, where the damping rate of each normal-mode depends on $x$. The resulting quantum noise interference implies that one could ground-state cool a mechanical resonator in the unresolved sideband-limit \cite{Elste2009}.  We also explain in detail how this quantum noise interference persists even outside the adiabatic limit, and show how standard approximations, which ignore noise correlations, can give misleading results.

Turning to measurement physics, we show that the interference-based noise cancellation can only be used to cancel all linear backaction effects in the unresolved sideband limit, where the cavity damping rate is greater than the mechanical frequency.  In contrast, one needs to be in the opposite (good-cavity) limit to make a QND measurement of mechanical energy \cite{Thompson2008,Jayich2008,Ludwig2012}.  
As a result, even in an ideal one-port device QND measurement of the mechanical energy that can resolve single-phonon jumps requires 
a single-photon optomechanical coupling that is larger than the cavity damping rate.  More optimistically, the cancellation of linear backaction in single-port system in the bad cavity limit allows one to make a measurement of $x^2$; such nonlinear measurement can lead to non-classical mechanical states \cite{Jacobs2009,Vanner2011}.


\section{Model}

\subsection{Hamiltonian and coupling to environment}

The two-cavity optomechanical system of interest is depicted in \cref{fig:diagram}. Its Hamiltonian takes the form
\begin{equation}
	\begin{split}
	\hat  H & = \hat  H_{\rm opt} +  \hat  H_{\kappa} + \hat  H_{\rm m} + \hat H_{\gamma} + \hat{H}_{\rm int}.
	\label{eq:Htot}
	\end{split}
\end{equation}
The Hamiltonian describing the cavities in the absence of optomechanical coupling is ($\hbar = 1$ throughout)
\begin{equation}\begin{split} 
	\hat  H_{\rm opt} 
		& =  \omega_{c}\p{\hat  a_{L}\dg\hat  a_{L} + \hat  a_{R}\dg\hat  a_{R}} - J\p{\hat  a_{L}\dg\hat  a_{R} + \hat  a_{R}\dg\hat  a_{L}}
		\\
		 & = \p{\omega_{c} - J}\hat a_{+}\dg\hat a_{+} + \p{\omega_{c} + J}\hat a_{-}\dg\hat a_{-}
\end{split}\end{equation}
where $\hat  a_{L}, \hat  a_{R}$ are the annihilation operators for a pair of localized cavity modes $L, R$ (e.g.~to the left and to the right of a membrane at the center of an optical cavity), and $J$ describes a tunnel coupling between them.  In the second line, we have diagonalized this Hamiltonian in terms of the optical normal modes $\hat  a_{\pm} = \tfrac{\hat  a_{L} \pm \hat  a_{R}}{\sqrt{2}}$.  
The mechanical resonator (frequency $\omega_{\rm m}$, annihilation operator $\hat{b}$, 
position operator $\hat{x}$) is described by $\hat{H}_{\rm m} = \omega_{\rm m} \hat{b}^\dagger \hat{b}$.  The optomechanical coupling takes the form 
$\hat{H}_{\rm int} = -( \hat{x} / x_{\rm zpt}) \hat{F}_{\rm opt}$ where
\begin{equation}
	\hat F_{\rm opt} 
	= g \left( \hat{a}^\dagger_L \hat{a}_L - \hat{a}^\dagger_R \hat{a}_R \right)
	 = g \p{\hat  a_{+}\dg\hat a_{-} + \hat a_{-}\dg\hat a_{+}}
\end{equation}
and $x_{\rm zpt}$ is the ground state position uncertainty of the oscillator.  Note that we have scaled the backaction force operator to have units of a rate.

The remaining terms in \cref{eq:Htot} describe dissipation.  As is standard, the mechanical dissipation ($\hat{H}_\gamma$) 
is described by a coupling to a thermal Markovian reservoir, giving rise to an amplitude damping rate $\gamma/2$.  The treatment of cavity dissipation (described by $\hat{H}_{\kappa}$) requires slightly more care.   We take each of the $L, R$ modes to be coupled to 
{\it independent}, zero-temperature Markovian reservoir.  The dissipation of the cavities is thus described by
\begin{equation}
	\hat H_{\kappa} = -i\left( \sqrt{\gk_{L}}\hat\Xi_{L}\dg\hat a_{L} + \sqrt{\gk_{R}}\hat\Xi_{R}\dg\hat a_{R} - \hc\right) + \hat{H}_{\rm ext}
	\label{eq:Hkappa}
\end{equation}
where $\hat \Xi_{L,R}$ are bath operators, and $\hat{H}_{\rm ext}$ describes the free bath modes.   
We treat the dissipation as per standard input output theory (see e.g. \cite{Gardiner2004,Clerk2010}). For $J=0$, the $L$ mode ($R$ mode) has
an amplitude damping rate $\kappa_L/2$ ($\kappa_R/2$).  Taking $L$ and $R$ to see independent dissipation describes dissipation a general,
asymmetric mebrane-in-the-middle type setup \cite{Thompson2008}, where each end mirror of the cavity has a non-zero transmission.  
We discuss the more general situation, where the two optical eigenmodes are generically coupled to two dissipative reservoirs (e.g.~a one port setup with internal loss) in \cref{app:gensetup}.

While \cref{eq:Hkappa} seems innocuous enough, it implies that unless $\kappa_R = \kappa_L$, the optical eigenmodes $\hat{a}_{\pm}$ {\it will not} be coupled to independent baths. This implies both that dissipation can couple the two optical normal modes, and that their fluctuations are correlated.  While it is standard in many quantum optics contexts to ignore such noise correlations between modes that are well-separated in frequency, we show they are significant in this class of systems, as the effects of non-resonant fluctuations are important.  Elucidating the consequences of the noise correlations our system when $\kappa_L \neq \kappa_R$ has not been fully done in previous work, and is the main goal of our work.    

\subsection{Heisenberg-Langevin equations}

To see this noise correlation explicitly, it is useful consider the Heisenberg-Langevin equations for our system in the absence
of optomechanical coupling.  We consider the standard case where the cavity modes are coherently driven at a frequency $\omega_{\rm dr}$ near the symmetric normal-mode frequency $\omega_c - J$.  Working in a rotating frame at the drive frequency, the optical Hamiltonian takes the form
\begin{equation}\begin{gathered}
\hat  H_{\rm opt} =  -\gd\hat a_{+}\dg\hat a_{+} + \p{2J - \gd}\hat a_{-}\dg\hat a_{-}
\end{gathered}\end{equation}
where $\gd = \omega_{\rm dr} - (\omega_c - J)$ is the detuning of the driving frequency from the $\hat{a}_{+}$ resonance.

By applying standard input-output theory \cite{Gardiner2004,Clerk2010}, the Heisenberg-Langevin equations of motion for our system (at $g=0$)
are easily found.  Defining ${\bar\gk = \half[\gk_{L}+\gk_{R}]}$, ${\Delta\gk = \half[\gk_{L}-\gk_{R}]}$, we have:
\begin{equation}
\begin{split}
	\dot{\hat a}_{-} & = -\p{\half[\bar\gk] + i\gd - 2iJ}\hat a_{-} - \half[\Delta\gk]\hat a_{+}
	\\ 
	&   
		+ \mat{ \sqrt{\half[\gk_{L}]} \ga_{L}^{\rm in}  - \sqrt{\half[\gk_{R}]} \ga_{R}^{\rm in} }
		+ \mat{ \sqrt{\half[\gk_{L}]} \hat\xi_{L} - \sqrt{\half[\gk_{R}]} \hat \xi_{R} },
	\label{eq:eomam}
\end{split}
\end{equation}
\begin{equation}
\begin{split}
	\dot{\hat a}_{+} & = -\p{\half[\bar\gk] + i\gd}\hat a_{+} - \half[\Delta\gk]\hat a_{-}
	\\ 
	&   
		+ \mat{ \sqrt{\half[\gk_{L}]} \ga_{L}^{\rm in}  + \sqrt{\half[\gk_{R}]} \ga_{R}^{\rm in} }
		+ \mat{ \sqrt{\half[\gk_{L}]} \hat\xi_{L} + \sqrt{\half[\gk_{R}]}  \hat \xi_{R} }.
\label{eq:eomap}
\end{split}\end{equation}
We are considering the general case where a coherent cavity drive is applied both at the left port and the right port, with respective classical input field amplitudes 
$\alpha^{\rm in}_L$ and $\alpha^{\rm in}_R$.  The $\hat \xi_i\p{t}$ operators ($i=L,R$)  describe operator-valued Gaussian white noise, i.e.~incident vacuum fluctuations entering the left and right port. They have zero mean and correlation functions
$\avg{\hat \xi_{i}\dg\p{t}\hat \xi_{j}\p{t\pr}} = 0$, $\avg{\hat \xi_{i}\p{t} \hat \xi_{j}\dg\p{t\pr} } = \gd_{ij}\gd\p{t-t\pr}$.

One sees clearly that the noise fluctuations driving $\hat{a}_+$ and $\hat{a}_{-}$ are in general correlated with one another.  These noises are of course identical (and hence completely correlated) in the case of a one-port cavity, $\kappa_R = 0$.  The only case where there is no correlation  is when $\kappa_R = \kappa_L$.  In this case, the situation is identical to having the $+$ and $-$ optical modes coupled to independent reservoirs.

\subsection{Effective dissipative optomechanical coupling}
\label{sec:kx}

For more intuition into the origin of linear backaction effects in this system, it is useful to consider the structure of the ``adiabatic" cavity eigenmodes:  if $\hat{x}$ is treated as a static, classical parameter, then what are the cavity eigenmodes for a given value of $x$?  One finds the lower-energy adiabatic eigenmode is described by 
\begin{equation}
	\hat{a}_{+}\br{x} = \cos\br{\theta\p{x}} \hat{a}_{L} +   \sin\br{\theta\p{x}} \hat{a}_{R},
\end{equation}
where $\cot2\theta\p{x} = gx/Jx_{\rm zpt}$. Its frequency is given by ${\omega_{+}\br{x} = \omega_{c} - \sqrt{J^2 + \p{g x / x_{\rm zpt}}^2}}$, and depends quadratically on $x$. 

Despite the $x^2$ dependence of the adiabatic mode frequencies, the dissipation of the adiabatic optical modes can lead to linear backaction. As it couples to the environment through the two leaky mirrors at $L, R$, the dissipation rate of the $+$ adiabatic mode is
\begin{align}
	\kappa_{+}[x] & = \cos^2 \br{\theta(x)} \kappa_L + \sin^2\br{\theta(x)} \kappa_R \nonumber \\
		& = 
		\bar{\kappa} + \frac{g x/x_{\rm zpt} } {\sqrt{J^2  + \p{g x/x_{\rm zpt}}^2}} \Delta \kappa.
\end{align}
Unless $\kappa_L = \kappa_R$, the effective damping rate of the adiabatic mode depends on whether the mode is localized more on the left or on the right.  As the ``wavefunction" of this adiabatic mode depends on $x$, one obtains a so-called dissipative optomechanical coupling \cite{Elste2009}, where the damping rate of an optical mode depends on $x$.  Note that this represents a potentially simpler method for realizing a dissipative optomechanical coupling than the Michelson-Sagnac interferometer proposed in Ref.~\onlinecite{Xuereb2011} and realized experimentally in Ref.~\onlinecite{Sawadsky2015}.  We note that recent experiments using trampoline-style resonators in a Fabry-Perot cavity with $\kappa_L \gg \kappa_R$ observed large position-dependent photonic damping \cite{Reinhardt2015}.

Crucially, to leading order in $1/J$, we see that even though the adiabatic mode frequency has a quadratic dependence on $x$, $\kappa_+[x]$ depends {\it linearly} on $x$ for small displacements. This implies that there will be information on $x$ available in the cavity output field, opening the door to unwanted linear backaction effects.  It also suggests that the unusual quantum noise physics of a dissipative optomechanical coupling will be relevant here, namely the possibility of Fano-style interference \cite{Miao2009,Elste2009}.

Note that throughout this discussion we have focused on the lower-energy adiabatic mode, $\hat a_{+}$. A full discussion must include the higher-energy adiabatic mode as well, as the two are in general coupled by dissipation. Such coupling effects only contribute at higher orders in $1/J$.

\section{Backaction quantum noise spectrum\label{sec:SFF}}


We now turn to the question of how noise correlations influence the linear optomechanical backaction on the mechanical resonator.  We focus on the standard case where the optomechanical coupling is sufficiently weak that optical backaction effects on the mechanical resonator can be fully understood using linear response theory.  This is equivalent to extracting backaction effects from the quantum noise spectrum of the optical force operator $\hat{F}_{\rm opt}$, evaluated to leading order in 
$g$ \cite{Marquardt2007,Marquardt2008,Clerk2010}. We further assume that $g$ is so weak that one only needs to consider the drive-enhanced optomechanical coupling, i.e.~one can linearize the $\hat{F}_{\rm opt}$ operator in the fluctuations of $\hat{a}_{\pm}$.  

\subsection{Noise spectrum and noise amplitudes}

The (unsymmetrized) quantum noise spectral density of $\hat{F}_{\rm opt}$ is defined as \cite{Marquardt2007,Marquardt2008,Clerk2010}
\begin{equation}
S_{FF}\br{\omega} = \int dt\; e^{i\omega t}\avg{\hat F_{\rm opt}\p{t}\hat F_{\rm opt}\p{0}}.
\end{equation}

As long as the features $S_{ FF}\br{\omega}$ are wider than the mechanical linewidth, the cavity can be understood as an effective thermal bath for the mechanics, with optomechanical damping rate $\Gamma$, and effective thermal occupation $\bar n_{\rm eff}$ given by
\begin{equation}\begin{gathered}
\Gamma = S_{ FF}\br{\omega_{m}} - S_{ FF}\br{-\omega_{m}}, 
\quad \Gamma \bar n_{\rm eff} = S_{ FF}\br{-\omega_{m}}.
\label{eq:optparams}
\end{gathered}\end{equation}
Here, $S_{FF}\br{+\omega_{m}}$ describes emission of energy from the mechanics into the cavity, or cooling processes, while $S_{FF}\br{-\omega_{m}}$ describes absorption of energy by the mechanics, or heating processes.

In our system, the leading terms in $\hat{F}_{\rm opt}$ (which are enhanced by the classical cavity drive) will be linear in the input nois operators.  
We can thus write it in terms of ``noise amplitudes" $\mathcal{A}_{L/R}[\omega]$ as:
\begin{equation}\begin{split}
	\hat F_{\rm opt}\br{\omega} = \sum_{i=L,R}\mathcal A_{i}\br{\omega}\hat \xi_{i}\br{\omega} + \mathcal A_{i}^{*}\br{-\omega}\hat \xi_{i}\dg\br{\omega} 
\label{eq:Foptform}
\end{split}\end{equation}
We have defined $\hat X\br{\omega} \equiv \int_{-\infty}^{\infty} e^{i\omega t}\hat X\p{t}dt$ for any operator $\hat X$ (implying $\hat X\dg\br{\omega} = \br{\hat X\br{-\omega}}\dg$).  As 
the input noise operators describe vacuum noise, one immediately finds 
\begin{equation}
	S_{FF}\br{\omega} = \abs{\mathcal A_{L}\br{\omega}}^{2} + \abs{\mathcal A_{R}\br{\omega}}^{2}
	\label{eq:SFFSimple}
\end{equation}
In addition to controlling the backaction noise spectral density, the noise amplitudes $\mathcal{A}_{i}[\omega]$ also directly determine how well one can make a linear measurement of position $x$ from the output light leaving the cavities; large linear backaction effects come hand in hand with large amounts of information on $x$ in the output field. This is shown explicitly in Appendix \ref{app:LinearMeas}.

The amplitudes $\mathcal{A}_i[\omega]$ are found by solving the Heisenberg-Langevin equations (\ref{eq:eomam})-(\ref{eq:eomap}) in the Fourier domain.  They are each a sum of two terms, corresponding to the two optical normal modes $\hat{a}_+$ and $\hat{a}_-$: 
\begin{equation}\begin{gathered}
	\mathcal A_{L/R}\br{\omega} = \tfrac{i}{\sqrt{2}}\tfrac{G}{2\tilde J}\sqrt{ \gk_{L/R}}\times
	\\ 
	\bmat{\tfrac{\gve_{m} + \p{i\half[\Delta\gk] \mp \Delta J}\p{1 \pm \tfrac{\gve_{m}}{2J}}}{\omega + \gd -\Delta J + i \half[\bar\gk]}
 	-  \tfrac{\mp2J + \p{ i \half[\Delta\gk]\pm\Delta J }\p{1 \pm \tfrac{\gve_{m}}{2J}}}{\omega + \gd - \Delta J - 2\tilde J + i \half[\bar\gk]}}
	\label{eq:FoptA}
\end{gathered}\end{equation}
where
\begin{equation}\begin{gathered}
G = g \abs{\avg{\hat a_{+}}} \quad \tfrac{\gve_{m}^{*}}{2J} = \tfrac{\avg{\hat a_{-}}}{\avg{\hat a_{+}}} 
\\ \tilde J = \sqrt{J^{2} - \p{\half[\Delta\gk]}^{2}} \quad \Delta J = J - \tilde J.
\end{gathered}\end{equation}
Here, $G$ is the many-photon optomechanical coupling, $\gve_m$ is the ratio of the average amplitudes of the optical eigenmodes, and $\Delta J$ is the correction to the eigenmode splitting frequency due to dissipation.

The form of $\mathcal{A}_{i}[\omega]$ has a simple interpretation: the vacuum noise entering each port of the cavity can contribute to the force noise in two ways, either by exciting the symmetric mode (the first term in the brackets of \cref{eq:FoptA}),  or the anti-symmetric mode (the second term in \cref{eq:FoptA}).  For small frequencies, the first process is near-resonant, while the second far from resonant.  From \cref{eq:FoptA,eq:SFFSimple}, we see that interference between these two ``paths" will in general be important to determining the final value of the noise spectrum.  In particular, we have the possibility of using destructive interference to strongly suppress the noise at a given frequency.  The fact that interference could play a role in the backaction noise in this system when $\kappa_R = 0$ was briefly mentioned by Miao et al.~\cite{Miao2009}.
The interference here is also reminiscent of the backaction cancellation approach used by by Caniard et al.~\cite{Caniard2007}, where two mechanical modes responded to the same fluctuating radiation pressure force.

\begin{figure}[htbp] 
   \centering
   \includegraphics[width=\columnwidth]{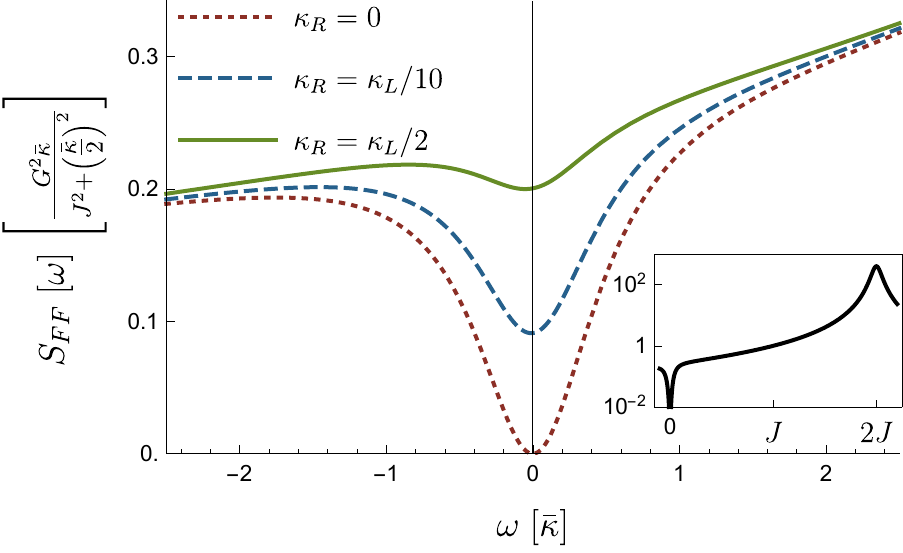} 
  
   \caption{The spectral noise density $S_{FF}\br{\omega}$ versus frequency $\omega$, for a two-mode optomechanical cavity driven from the left at the $+$ mode resonance 
   (i.e.~$\gd=0$).  We set $J = 10\bar\gk$ to be in the adiabatic regime; curves correspond to different values of $\kappa_R$. 
   At $\omega = 0$, the noise density is suppressed as $\gk_{R}/\bar\gk$. For ${\omega \gtrsim \bar\gk}$, the noise density returns to a near-constant value. Inset: $S_{FF}[\omega]$, but now plotted over a wider range of frequencies.  One clearly sees two resonances (one near $\omega=0$, one near $\omega \sim 2J$), corresponding to the two optical normal modes. At this scale all three curves overlap.}
   \label{fig:SFF}
\end{figure}

\subsection{Large-$J$ ``adiabatic" limit}
\label{subsec:LargeJ}

We now specialize to the usual situation where the normal-mode splitting $J$ is large, taking ${\abs{\omega}, \abs{\delta}, \bar\gk \ll 2J}$. The amplitudes determining the force noise then simplify, to leading order in $1/J$, to
\begin{equation}\begin{gathered}
	\mathcal A_{L/R}\br{\omega} = \tfrac{i}{\sqrt{2}}\tfrac{G}{2 J}\sqrt{ \gk_{L/R}}
	 \bmat{\tfrac{-\gd + i\half[\bar\gk]}{\omega + \gd + i \half[\bar\gk]}\gL^{*} \mp 1}
	\label{eq:FoptlJ}
\end{gathered}\end{equation}
where $\gL = \tfrac{\sqrt{\gk_{L}}\ga_{L}^{in} - \sqrt{\gk_{R}}\ga_{R}^{in}}{\sqrt{\gk_{L}}\ga_{L}^{in} + \sqrt{\gk_{R}}\ga_{R}^{in}}$. 

In this regime, the non-resonant term in each amplitude reduces to a frequency-independent constant.  The resulting interference implies that the quantum noise spectral density as a function of $\omega$ (as given by \cref{eq:SFFSimple}) is the sum of two Fano lineshapes \cite{Fano1961,Miroshnichenko2010}.  In general, Fano lineshapes can exhibit perfect destructive interference with a vanishing net amplitude.  In our case, this will not be possible for both $\mathcal A_{L}$ and $\mathcal A_{R}$ simultaneously, because of the sign difference in the second term in \cref{eq:FoptlJ}.  Hence, at best interference can be used to cancel the noise coming from one port, at one particular frequency. 

The above destructive interference becomes even more explicit when one looks at the spectral density, which to leading order in $1/J$ is
\begin{align}
	S_{ FF}\br{\omega} & = \tfrac{G^{2}}{4J^{2}}\bar\gk\times 
	\nonumber \\
	& \bmat{ \frac{\gk_{L}\gk_{R}}{\bar\gk^{2}} + \tfrac{\abs{\tfrac{\Delta\gk}{\bar\gk}\p{\omega + 2\gd} + \p{\gd + i\half[\bar\gk]}\p{\gL - \tfrac{\Delta\gk}{\bar\gk}}}^{2}}{\abs{\omega + \gd + i\half[\bar\gk]}^{2}}}.
\label{eq:SFF}
\end{align}
By setting the drive detuning $\delta$ (as well as the relative amplitude of the drives applied to each port) one can suppress the second term at a particular frequency, minimizing the noise at this frequency.  This is a direct manifestation of the destructive interference discussed above.  In particular, for backaction cooling applications, one could chose to minimize the noise at $\omega = -\omega_m$, as this minimizes $\bar{n}_{\rm eff}$ (c.f. \cref{eq:optparams}).  By using a drive detuning $\gd = \omega_{m}/2$ and assuming that we only drive the cavity from the $L$ port (i.e.~$\alpha^{\rm in}_{R}=0$), we find to leading order in $1/J$,
\begin{equation}
	S_{FF}[-\omega_M] = \frac{G^2}{2J^{2}} \kappa_R
	\label{eq:minSFF}
\end{equation}

\Cref{eq:minSFF} implies that the ``heating" backaction noise vanishes completely in the limit of a one port cavity (i.e. $\kappa_R \rightarrow 0$), so that the cavity backaction acts like a zero-temperature reservoir for the mechanics.  This opens the door to ground-state cooling of mechanical resonators that are not in the good cavity limit, something that cannot be done in a standard, coherently-driven single-cavity optomechanical system. We stress that in the large-$J$ limit, this quantum noise cancellation and cooling can be completely understood 
in terms of the effective dissipative optomechanical coupling in our system (as introduced in Sec.~\ref{sec:kx}).

Shown in \cref{fig:SFF} are representative plots of $S_{FF}[\omega]$ which illustrate the noise cancellation effect.  
The width in frequency of the noise suppression is $\Delta\omega \sim \bar\gk$. This implies that in the extreme bad-cavity limit $\omega_{m}\ll \bar\gk$,  one can use this interference to suppress {\it all} linear backaction effects, i.e. both heating and cooling.  In contrast, in the good cavity limit $\omega_{m} \gtrsim \bar\gk$, it is not possible to use interference to suppress both $S_{FF}[\omega_m]$ and 
$S_{FF}[-\omega_m]$.  As a result, linear backaction effects persist even in a pure single-port system where $\gk_{R}=0$.  As we discuss below, this means QND measurement of phonon number in a one-port cavity of this type is subject to the same tough constraints on the single-photon optomechanical coupling $g$ as in a two-port cavity.

\subsection{Noise interference away from the adiabatic, large-$J$ limit}

\begin{figure}[thbp] 
   \centering
   \includegraphics[width=\columnwidth]{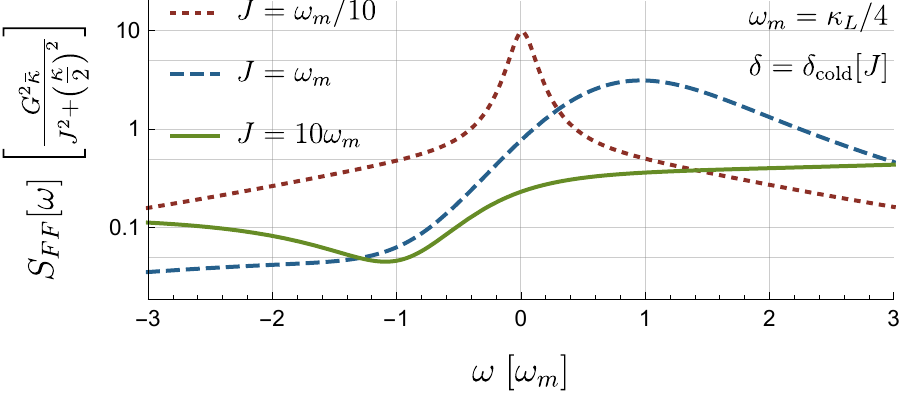} 
   
   \includegraphics[width=\columnwidth]{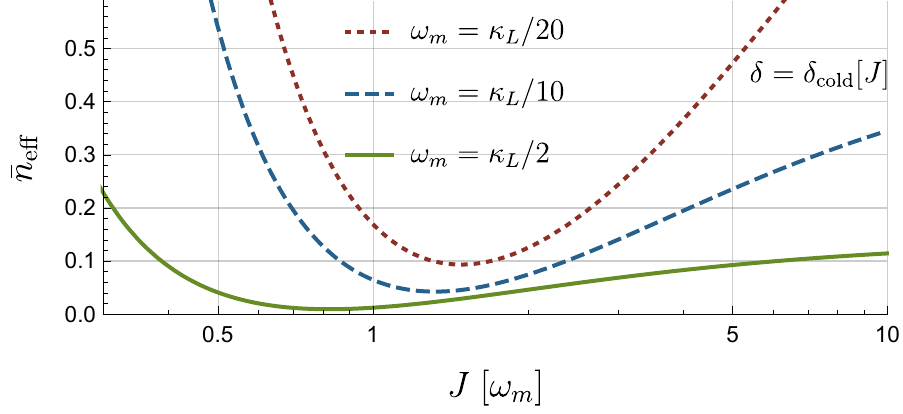} 

   \caption{Behaviour of the backaction noise spectral density $S_{FF}[\omega]$ away from the adiabatic, large-$J$ limit.  
   All results correspond to  $\gk_{R} = \gk_{L}/20$ and having set the drive detuning $\delta$ to its optimal value for for minimizing $S_{FF}\br{-\omega_{m}}$:   
   ${\gd = \delta_{\rm cold} = \half[\omega_{m}] + J - \sqrt{J^{2} + \p{\half[\omega_{m}]}^{2}}}$.
   Top: $S_{FF}[\omega]$ versus $\omega$, for different values of $J$.  All curves correspond to $\omega_{m} = \tfrac{\gk_{L}}{4}$. We observe the crossover behavior from a Fano-line shape to a single peak as $J$ is reduced. 
      Bottom: The effective thermal occupation $\bar n_{\rm eff}$ associated with the cavity backaction as a function of 
   the splitting $J$ of the optical normal modes.  It is minimized at the crossover regime $J\sim \omega_{m}$.
   }
   \label{fig:smallJ}
\end{figure}

While the noise spectrum is easiest to understand in the large-$J$ limit, we find that most of the features described above appear for any value of the mode splitting.
Away from the large-$J$ limit, $S_{FF}[\omega]$ does not exhibit a simple Fano resonance, but reflects the interference of the two resonant amplitudes written in \cref{eq:FoptA}.
As shown in \cref{fig:smallJ}, one finds a qualitative crossover in the form of the spectrum as $J$ is reduced below $\sim \gk$ (i.e.~when the cavity  normal modes are no longer resolved).

Despite the changes in the shape of $S_{FF}[\omega]$, we find that a perfect destructive interference of the noise is possible in the single port limit for \emph{any} value of $J$.  We stress that for small $J$, one is not in the adiabatic limit, and the system is not equivalent to having a dissipative optomechanical coupling.

For $\gk_{R} = 0$, the noise spectral density is
\begin{equation}
	S_{FF}\br{\omega} = \tfrac{2G^{2}\gk_{L}}{\p{2J-\gd}^{2}}\abs{\tfrac{J\p{\omega + 2\gd} - \gd\p{\omega + \gd}}{2J\p{\omega + \gd + i\tfrac{\gk_{L}}{4}} - \p{\omega + \gd}	\p{\omega + \gd + i\half[\gk_{L}]}}}^{2}.
\label{eq:SFFkr0}
\end{equation}
It can always be made to vanish at $S_{FF}\br{-\omega_{m}}$ by setting the detuning $\gd = \half[\omega_{m}]+J-\sqrt{J^{2} + \p{\half[\omega_{m}]}^{2}} \equiv \gd_{\rm cold}$. 
It might seem surprising that the interference persists even for very small values of $J$, as one would expect to recover the physics of a standard, single-cavity optomechanical system.  This is not the case.  On the dissipation-free side of the 
cavity, the average amplitude is inversely proportional $J$: for $\gk_R = 0$, 
$\avg{\hat a_{R}} = \tfrac{J - \gd}{J}\avg{\hat a_{L}}$. As long as there is no loss through the $R$ port, the amplitude in the right cavity can grow arbitrarily large, allowing for perfect destructive interference. The same intuition holds as long as $\gk_{R}$ remains small compared with $\omega_{m}$ and $J$; see \cref{sec:cooling} for a more quantitative analysis of this restriction.

 For $\gk_{R}>0$, the force noise cannot be made to vanish at any frequency, but the effective thermal occupancy $\bar{n}_{\rm eff}$ associated with the backaction can be minimized by choosing an appropriate drive detuning $\delta$.  In general, to achieve a small $\bar{n}_{\rm eff}$, one would like to both minimize $S_{FF}[-\omega_m]$ (i.e.~the ``heating" noise), while simultaneously maximizing $S_{FF}[\omega_m]$ (the ``cooling" noise).  The resonant structure of the noise spectrum has its minimum and maximum near its two poles, which are roughly $2J$ apart (see \cref{eq:FoptA} and the inset of \cref{fig:SFF}). This means that that $\bar n_{\rm eff}$ is minimized for $J\sim \omega_{m}$.


\section{Consequences for QND phonon measurement\label{sec:QND}}

\begin{figure}[t] 
   \centering
   \includegraphics[width=\columnwidth]{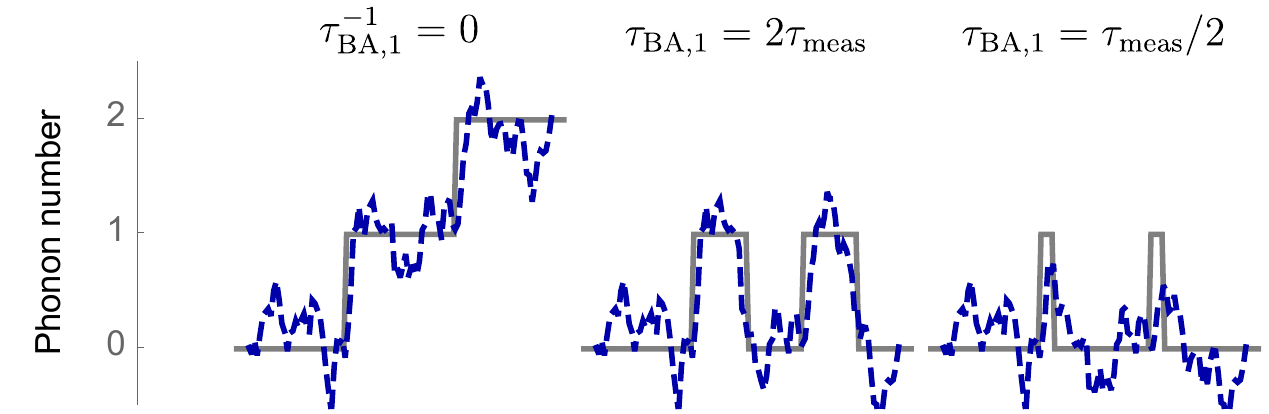} 
   \caption{Illustration of a QND measurement of the phonon number. Plotted as a function of time, the grey lines indicate the behaviour of the
   actual mechanical phonon number, while the dashed blue lines indicate the output signal, averaged over $\tau_{\rm meas}$. At times $t = 4\tau_{\rm meas}, 8\tau_{\rm meas}$ the mechanical system experiences a upward quantum jump due to thermal noise unrelated to the cavity. In the absence of backaction ($\tau_{{\rm BA},1}^{-1}=0$), or when the the typical backaction is longer than the measurement time ($\tau_{{\rm BA},0}=2\tau_{\rm meas}$), these jumps and the discreteness of the phonon numbers can be observed. When the backaction dominates ($\tau_{{\rm BA},1} = \tau_{\rm meas}/2$) the mechanical oscillator appears to remain in the ground state.}
   \label{fig:QJ}
\end{figure}

A key motivation for the study of membrane-in-the-middle style optomechanical setups is the possibility of quantum non-demolition (QND) measurement of mechanical energy eigenstates, and the possibility to observe ``quantum jumps" in the mechanical energy  \cite{Thompson2008,Jayich2008,Ludwig2012,Paraiso2015}.  As sketched in Sec.~\ref{sec:kx}
(and derived more rigorously in \cite{Thompson2008,Jayich2008}), in the large-$J$ limit one can adiabatically eliminate the off-resonant mode to obtain an effective optomechanical coupling
\begin{equation}
	\hat{H}_{\rm quad} = -\frac{g^2}{2J} \hat{a}^\dagger_{+} \hat{a}_{+} \left(\hat{b} + \hat{b}^\dagger \right)^2.
	\label{eq:Hquad}
\end{equation}
If one further assumes the good-cavity limit, the $\hat{b} \hat{b}$ and $\hat{b}^\dagger \hat{b}^\dagger$ terms have negligible influence, leaving only the desired coupling: the frequency of the cavity $+$ mode is controlled by the number of phonons in the mechanical resonator.  By driving the $+$ optical mode resonantly (i.e.~$\delta = 0$) and making a homodyne measurement of the optical phase quadrature, one can thus monitor the mechanical phonon number.  

\subsection{Measurement and backaction time scales}

Because of the intrinsic noise in the measured homodyne current, it will take a finite amount of time to resolve the mechanical phonon number.  A standard argument \cite{Miao2009} shows that the time needed to resolve the mechanical energy to better than one quanta is:
\begin{equation}
	\tau_{\rm meas} \sim \frac{ J^{2}\gk_{L}}{G^{2} g^{2} }
	\label{eq:TauMeas}
\end{equation}

As discussed previously \cite{Thompson2008,Jayich2008,Miao2009}, a successful QND measurement requires that linear backaction effects do not cause a transition of the mechanical state before the measurement can resolve it.  For a situation where the mechanical resonator is prepared near its ground state, a minimal requirement is that the measurement time $\tau_{\rm meas}$ be shorter than the lifetime of {\it both} the mechanical ground state and the $n=1$ Fock state due to backaction. This is illustrated in \cref{fig:QJ}.

\begin{figure}[t] 
   \centering
   \includegraphics[width=\columnwidth]{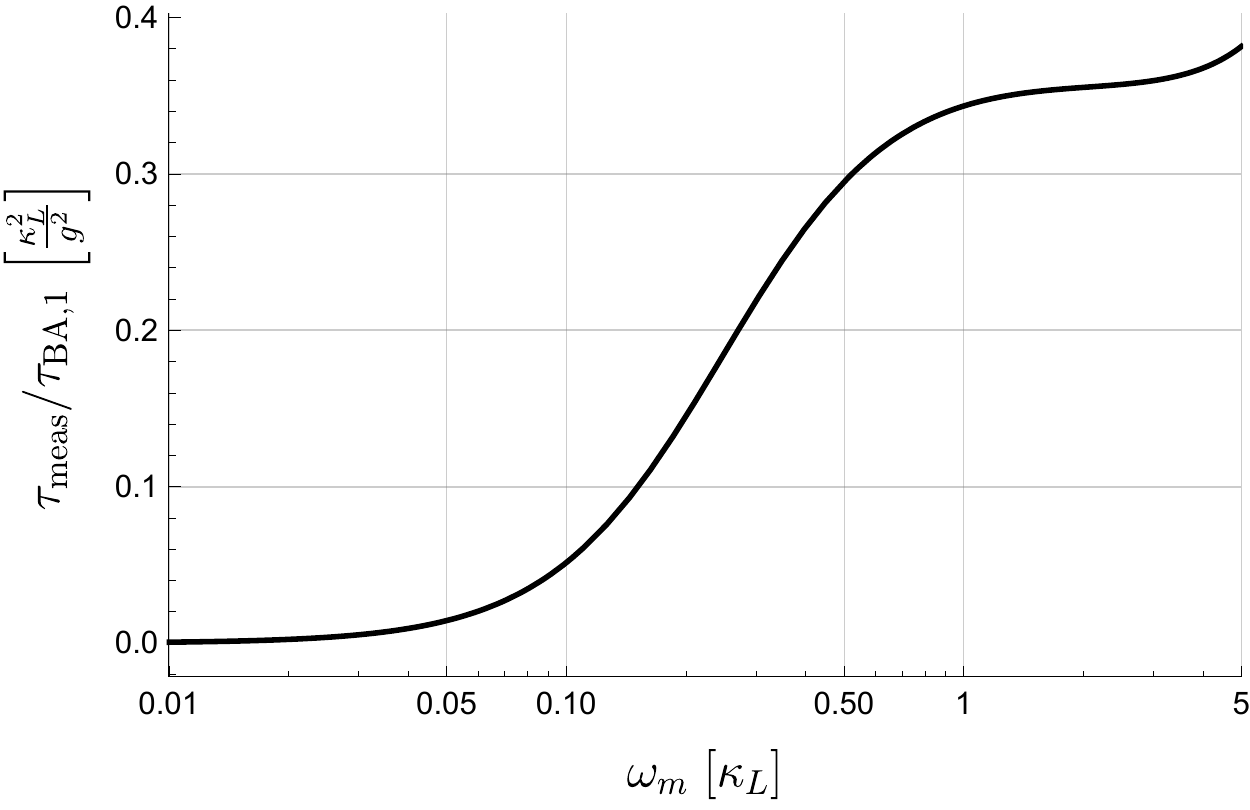} 
   \caption{The relative ratio between typical times for quadratic measurement and for linear backaction,  $\tau_{\rm meas}/\tau_{BA,1}$ (see \cref{eq:TauMeas,eq:linnoise,eq:linnoisen}). As discussed in the text, to perform QND measurements, this ratio must be less than one.
   Shown here for for a one-port cavity, $\gk_{R} = 0$, with $J = 10\gk_{L}$, $\gd = 0$, driving from the left.   At $\omega_{m}\ll \gk_{L}$, linear noise is strongly suppressed, while in the good cavity limit its value is similar to the two-port case.}
   \label{fig:linnoise}
\end{figure}

Fermi's Golden rule lets us directly relate the lifetime of the $n$th mechanical Fock state to the backaction quantum noise spectrum,
\begin{equation}\begin{split}
	\tau_{{\rm BA},n}^{-1} & \equiv (1+n)S_{\rm FF} \br{-\omega_m} + n S_{\rm FF} \br{+\omega_m}.
	\label{eq:linnoise}
\end{split}\end{equation}
For driving through the $L$ port at $\delta = 0$ we find
\begin{equation}\begin{split}
	\tau_{{\rm BA},n}^{-1} &  = \frac{G^{2}}{J^{2}}\br{\tfrac{\half[\bar\gk]\omega_{m}^{2} + \gk_{R}\p{\half[\bar\gk]}^{2}}{\omega_{m}^{2} + \p{\half[\bar\gk]}^{2}} + O\p{\tfrac{1}{J}}}\p{n + \half}.
	\label{eq:linnoisen}
\end{split}\end{equation}
As the mean backaction time decreases with $n$, the requirement for effective QND measurements is $\tau_{\rm meas}/\tau_{{\rm BA},1}<1$. This ratio is plotted, for a one-port cavity, in \cref{fig:linnoise}. 

For a completely symmetric two port cavity ($\kappa_L = \kappa_R = \bar\kappa$), \cref{eq:linnoisen} yields $\tau_{{\rm BA},1}^{-1} \sim \tfrac{G^{2}}{J^{2}} \bar\kappa$ (with a prefactor ranging from $\half$ to $1$). 
Combined with \cref{eq:TauMeas},  the requirement $\tau_{{\rm BA},1} > \tau_{\rm meas}$ for QND measurement then reduces to $g \gtrsim \bar\kappa$: the single-photon optomechanical coupling rate must be larger than the cavity damping rate.  This is in agreement with previous work \cite{Miao2009}. 

For an asymmetric cavity, we must consider separately the resolved and unresolved sideband limits.

\subsection{Resolved sideband limit}

To see the discreteness of the mechanical energy, one needs to be in the resolved-sideband (i.e.~good cavity) limit, 
$\omega_{ m} \gg \bar\gk$. In this regime $\hat{x}^2$ is approximately proportional to the phonon number operator, as discussed following \cref{eq:Hquad}. In the resolved-sideband limit, \cref{eq:linnoisen} indicates  $\tau_{{\rm BA},1}^{-1}$ scales as $\tfrac{G^{2}}{J^{2}}\bar\kappa$, i.e.~in the same way as in the symmetric two-port case.  The requirement that $\tau_{{\rm BA},1} > \tau_{\rm meas}$ thus reduces again to requiring $g \gtrsim \bar\gk$ for QND measurement. This means that the scale of the optomechanical coupling must be larger than both $\kappa_L$ and $\kappa_R$ (and not just their product).  A similar conclusion holds in general for a single-sided cavity having internal loss:  $g$ must be larger than the coupling-$\kappa$, not just the internal-loss $\kappa$ (see \cref{app:gensetup}).

The above conclusion differs from previous works, which suggested that in a perfect one-port cavity, linear backaction effects do not present any limit to performing QND measurement.  While it is true that in a single port cavity, one can perfectly cancel the backaction noise at $\omega = -\omega_m$ via interference (c.f. \cref{eq:SFF}), the noise at positive frequency $\omega = + \omega_{\rm m}$ remains.  This noise will kill the lifetime of the $n=1$ Fock state, making it impossible to resolve a quantum jump (see \cref{fig:QJ}).  

Finally, we note that if one manages to achieve a very different situation than that described here, where
each optical normal mode couples independently to a separate dissipative reservoir, then one recovers the result of Ref.~\onlinecite{Miao2009}:  
the condition $\tau_{{\rm BA},1} > \tau_{\rm meas}$ reduces to 
$g^2 \gtrsim \kappa_{+} \kappa_{-}$.

\subsection{Unresolved sideband limit and possibility of $x^2$ measurement}

In the $\omega_{m}\ll \bar\gk$ limit, we have already shown that noise interference can be used to completely cancel linear backaction effects, 
see Sec.~\ref{subsec:LargeJ}.  Thus, not surprisingly, in this limit we find that $\tau_{{\rm BA},1}$ diverges as $\gk_{R} \rightarrow 0$. In the perfect one-port case, $\gk_{R} = 0$, there is no linear backaction at all.

In this limit Eq.~(\ref{eq:Hquad}) does not allow the cavity to measure phonon number.  Instead, the cavity will measure $\hat x^2$ of the mechanics. While such a measurement
does not allow one to detect quantum jumps in mechanical energy, its nonlinear nature can allow the conditional generation of highly non-classical mechanical states, i.e. states which exhibit negativity in their Wigner functions \cite{Jacobs2009,Vanner2011}.  We will explore this physics in detail in a later work.

\section{Backaction cooling}
\label{sec:cooling}

In this final section, we return to quantum noise interference that is possible in our system, and discuss further the possibilities for mechanical cooling.  In the large-$J$, adiabatic limit, and for $\kappa_R = 0$, the fact that our system allows ground-state cooling of a mechanical resonator in the bad-cavity limit is not surprising, as it directly realizes the dissipative $\kappa(x)$ coupling discussed in Ref.~\onlinecite{Elste2009} (see \cref{sec:kx}).  More surprising is the fact that ground state cooling is possible even away from the large-$J$ limit, where the system is not identical to the dissipative-coupling optomechanical system.   As we have seen in \cref{eq:SFFkr0} and the surrounding discussion, an effective zero temperature can be achieved for any $J$ given $\gk_{R} = 0$.

Driving through the $L$ port, setting $\gd = \gd_{\rm cold}$ (see \cref{eq:SFFkr0} and the following discussion), we find to leading order in $\gk_{R}$,
\begin{align} 
	\Gamma & = \tfrac{2G^{2}}{J^{2}} \tfrac{\gd_{\rm cold}^2}{\omega_m^2 \p{\bar n_{\rm eff}/\gk_{R}}} \\
	\bar n_{\rm eff} & = \tfrac{9}{4}\p{\sqrt{1 + \p{\tfrac{\omega_{m}}{2J}}^{2}} - \tfrac{5}{3}\tfrac{\omega_{m}}{2J}}^{2}\tfrac{\gk_{R}}{\gk_{L}}
	\nonumber \\ 
	& \quad  + \p{\sqrt{1 + \p{\tfrac{\omega_{m}}{2J}}^{2}} - 3\tfrac{\omega_{m}}{2J}}^{2}\tfrac{\gk_{L}\gk_{R}}{16\omega_{m}^{2}}
\label{eq:optpar}
\end{align}
The system can be effectively cooled as long as the internal damping is small enough that $\gk_{L}\gk_{R}\ll \omega_{m}^{2}$. This range of parameters is experimentally realistic \cite{Reinhardt2015}.  We stress that this small $\kappa_R$ expansion does not assume anything about the value of $J$.  We also remind the reader that this result is derived within the perturbative, quantum-noise approach.  For strong drives, where $\Gamma\sim\bar\gk$, the broadening of the mechanical resonance by $\Gamma$ leads to an additional non-zero term in 
$\bar{n}_{\rm eff}$ \cite{Weiss2013,Weiss2013a}.  

Finally, note that in the good cavity limit, $\omega_{m}\gtrsim \gk_{L}$, more effective cooling is possible regardless of $\gk_{R}$ by setting the detuning to $\gd \sim -\omega_{m}$, using the same physics seen in a single-mode cavity \cite{Marquardt2008}.

\section{Conclusions and outlook}

We have presented a thorough analysis of the residual linear backaction noise in an asymmetric two-mode optomechanical system having the
form of the canonical ``membrane-in-the-middle'' system.  Our analysis shows that in the adiabatic, large-$J$ limit, the system has an effective dissipative optomechanical coupling, with a corresponding Fano interference in its quantum backaction noise.  Our analysis also shows that this interference (and potential for perfect cancellation) persists even to the non-adiabatic regime for arbitrary $J$.  We demonstrated that in a perfect one-port device in the unresolved sideband regime, all linear backaction effects could be suppressed, allowing an ideal continuous measurement of $x^2$.  In contrast, if one works in the good cavity limit and attempts to measure quantum jumps in the mechanical phonon number, linear backaction cannot be completely suppressed even in a single-port system, and one requires $g\gtrsim \bar \gk$ for the measurement to be stronger than the unwanted backaction.  

We have focused here primarily on a literal membrane in the middle style cavity, but our results hold, qualitatively, for other optomechanical setups where two optical modes are coupled to the mechanics (see \cref{app:gensetup}).  The only exception is a system where the two relevant optical normal modes see completely independent dissipation; such systems may be the most promising avenue for the observation of mechanical quantum jumps.  An interesting future direction would be to understand how quantum noise interference effects of the sort discussed here manifest themselves in even more complex multi-mode optomechanical systems.  

\section*{Acknowledgements}

This work was supported by NSERC.

\appendix

\section{Connection to effective linear measurement}
\label{app:LinearMeas}

We show briefly here that the noise amplitudes $\mathcal{A}_{i}[\omega]$ introduced in Eq.~(\ref{eq:Foptform}) control how well one could make a measurement of linear position from the output light leaving the cavity.  From standard input-output theory, the output field for port $i = L,R$ is given by \cite{Clerk2010}:
\begin{align}
	\hat a_{i}^{\rm out} = \hat a_{i}^{\rm in} - \sqrt{\gk_{i}} \hat a_{i}
\end{align}	

Using our Heisenberg-Langevin equations, one straightforwardly finds that the fluctuating part of the output field is given by
\begin{equation}
	\hat \xi^{\rm out}_{i}\br{\omega} = 
		\sum_{j=L,R} \mathcal B_{ij}\br{\omega}\hat \xi_{j}\br{\omega} - i\mathcal A_i[\omega] \hat x[\omega].
\end{equation}
Here, the $ \mathcal B_{ij}\br{\omega}$ are complex functions of frequency which are independent of the optomechanical coupling $g$; they determine the output fluctuations in the absence of any coupling. The second term describes how the output fields depend linearly on $\hat{x}$; note that the Heisenberg operator $\hat{x}$ here includes the effects of backaction.  We thus see that the linear-response kernel linking the output field to $\hat{x}[\omega]$ is identical to the noise amplitudes $\mathcal{A}_i[\omega]$.  This result also follows from standard quantum linear response theory (i.e.~the Kubo formula), which says that the relevant linear response kernel is given by the commutator of $\xi^{\rm out}_i(t)$ and the backaction force operator $\hat{F}(t)$ (see, e.g., Ref.~\onlinecite{Clerk2010}).  

It is illustrative to examine this output in the time domain where we can write, 
\begin{equation}
	\Delta\hat \xi_{L} \p{t} = \int_{-\infty}^{t}\mathcal G_{}\p{t-t'}\hat{x}(t') d t'
\end{equation}
for $\Delta\hat\xi_{L}\p{t}$, the portion of the output due to coupling to the mechanics. The response function is given by, to first order in $1/J$, by
\begin{equation}
	\mathcal G\p{\tau} = \tfrac{G}{2J} 
			\sqrt{\half[\gk_{L}]}
			\br{\half[\bar\gk] e^{-\half[\bar\gk] \tau} - \gd \p{\tau - \eta }}.
\label{eq:GAA}
\end{equation}
having taken $\gL = 1, \gd = 0$. Here, the first terms results from the resonant piece of $\mathcal A_{L}\br{\omega}$, while the second term comes from the off-resonant piece; $\eta$ is a positive infinitesimal.  

We can see in \cref{eq:GAA}  the two regimes described in \cref{sec:QND}, caused by the different response rates of the two cavity modes. The response of the rapidly oscillating~$-$ mode is near immediate. In the bad cavity regime, when $\omega_{m}\ll \bar\gk$, the response of the $+$ mode is faster than the rate of change in $\hat x$, leading to a perfect destructive interference. In the good cavity regime, $\omega_{m}\gg \bar\gk$, the slow response of the $+$ mode means $\hat x$ information is averaged out. However, this still leaves the information leaking through the $-$ mode, which is no longer canceled.

\section{Generic setup\label{app:gensetup}}

For completeness, we now discuss the generic case, the the cavity is coupled to its environment by some set of dissipation channels. The damping hamiltonian of any system coupled to a single driven port and any number of additional internal loss channels can be written in the form
\begin{equation}\begin{split}
\hat H_{\rm damp} &= -i\hat \Xi_{\rm dr}\dg\p{\sqrt{\gk^{+}_{\rm dr}} \hat a_{+} + \sqrt{\gk^{-}_{\rm dr}} \hat a_{-}}
\\ & \quad -i\hat \Xi_{\rm int}\dg\p{\sqrt{\gk^{+}_{\rm int}} \hat a_{+} - \sqrt{\gk^{-}_{\rm int}}\hat a_{-}} + \hc
\end{split}\end{equation}
where $\hat \Xi_{\rm dr},\Xi_{\rm int}$ are two independent linear combinations of the various dissipation modes, defined so that there is no driving through $\hat \Xi_{\rm int}$. We define here
\begin{equation}\begin{gathered}
\gk_{+} = \gk^{+}_{\rm dr} + \gk^{+}_{\rm int} \qquad \gk_{-} = \gk^{-}_{\rm dr} + \gk^{-}_{\rm int}
\\ \gk_{\rm dr} = \gk^{+}_{\rm dr} + \gk^{-}_{\rm dr} \qquad \gk_{\rm int} = \gk^{+}_{\rm int} + \gk^{-}_{\rm int}
\\ \bar\gk = \half[\gk_{\rm dr} + \gk_{\rm int}] \qquad \gd\gk = \half[\gk_{+} - \gk_{-}]
\\ \Delta\gk  = \sqrt{\gk_{\rm dr}^{+}\gk_{\rm dr}^{-}} - \sqrt{\gk_{\rm int}^{+}\gk_{\rm int}^{-}}
\end{gathered}\end{equation}

At $g = 0$, the equations of motion are
\begin{equation}\begin{split}
\dot{\hat a}_{-} & = -\p{\half[\gk_{-}] + i\p{2J-\gd}}\hat a_{-} - \half[\Delta\gk]\hat a_{+}
\\ & \quad  + \sqrt{\gk_{\rm dr}^{-}}\p{\ga^{in} + \hat \xi_{\rm dr}} - \sqrt{\gk_{\rm int}^{-}}\hat \xi_{\rm int}
\\ \dot{\hat a}_{+} & = -\p{\half[\gk_{+}] - i\gd}\hat a_{+} - \half[\Delta\gk]\hat a_{-}
\\ & \quad  + \sqrt{\gk_{\rm dr}^{+}}\p{\ga^{in} + \hat \xi_{\rm dr}} + \sqrt{\gk_{\rm int}^{+}}\hat \xi_{\rm int}
\end{split}\end{equation}
and we find
\begin{equation}
\frac{\gve_{m}^{*}}{2J} = \frac{\avg{\hat a_{-}}}{\avg{\hat a_{+}}} = \frac{\gd + i \half[\gk_{+}] - i\half[\Delta\gk]/t_{d}}
{-2J+\gd + i\half[\gk_{-}] - i \half[\Delta\gk]t_{d}}t_{d}
\end{equation}
and 
\begin{equation}\begin{gathered}
	\mathcal A_{\rm dr/int}\br{\omega} = \tfrac{iG}{2\tilde J}\sqrt{\gk_{\rm dr/int}^{+}}\times
	\\ \begin{split}
	\Bigg[&\tfrac{\gve_{m}\p{1 + \tfrac{i\gd\gk}{2J}} + i\half[\Delta\gk]\p{1 \pm \tfrac{\gve_{m}}{2J}t_{d/i}}
	\mp \Delta J\p{t_{d/i} \pm \tfrac{\gve_{m}}{2J}}}{\omega + \gd -\Delta J + i \half[\gk_{+}]} - 
	\\ &  \tfrac{\mp2Jt_{d/i}\p{1 + \tfrac{i\gd\gk}{2J}} + i\half[\Delta\gk]\p{1 \pm \tfrac{\gve_{m}}{2J}t_{d/i}}
	\mp \Delta J\p{t_{d/i} \pm \tfrac{\gve_{m}}{2J}}}{\omega + \gd - \Delta J - 2\tilde J + i \half[\gk_{-}]}\Bigg]
	\end{split}
\end{gathered}\end{equation}
where $t_{d/i} = \sqrt{\gk_{dr/int}^{-}/\gk_{dr/int}^{+}}$ and here 
\begin{equation}
\tilde J = \sqrt{\p{J+i \half[\gd\gk]}^{2} - \p{\half[\Delta\gk]}^{2}} \qquad \Delta J = J+i \half[\gd\gk] - \tilde J.
\end{equation}

For a small internal loss, the noise spectral density becomes, to leading order in $1/J$,
\begin{equation}\begin{gathered}
S_{\rm FF} = \tfrac{G^{2}}{4J^{2}}\tfrac{1}{\abs{\omega + \gd + i\half[\gk_{+}]}^{2}}\times
\\ 
\bmat{\gk_{\rm dr}^{-}\abs{\omega + 2\gd}^{2}
 +  \gk_{\rm int}^{-}\abs{\omega + \gd\p{1 - \tfrac{t_{d}}{t_{i}}} + i\half[\gk_{+}]\p{1 + \tfrac{t_{d}}{t_{i}}}}^{2}
}.
\label{eq:SFFgen}
\end{gathered}\end{equation}

We see that while the factors vary, the structure of the force noise and noise density spectrum are similar to those discussed in the main text, and seen in \cref{eq:FoptA,eq:SFF}. 

In particular, for the purpose of QND measurement, a single-port system in the good cavity limit has ${\tau_{{\rm BA},1} \sim \tfrac{G^{2}}{J^{2}}\gk_{\rm dr}^{-}}$, and leading to the requirement $g\gtrsim \gk_{\rm dr}^{-}$. Thus, QND measurements are possible when the off-resonant channel is not coupled to the driving port.

\bibliographystyle{apsrev4-1}
\bibliography{/Users/yarivyanay/Documents/University/Citations/library}

\end{document}